\documentclass[10pt, conference]{IEEEtran}

\usepackage{amsfonts}
\usepackage{mathrsfs}
\usepackage{graphicx}
\usepackage{cite}
\usepackage{citesort}
\usepackage{color}
\usepackage{psfrag}
\usepackage{subfigure}
\usepackage{amssymb}
\usepackage{epsfig}
\usepackage{pifont}
\usepackage{amsmath}
\usepackage[amsmath,thmmarks]{ntheorem}
\usepackage{array}
\usepackage{multicol}
\usepackage{algorithm}
\usepackage{algorithmic}
\allowdisplaybreaks[1]

\DeclareMathOperator*{\argmin}{arg\,min}
\DeclareMathOperator*{\argmax}{arg\,max}

\renewcommand{\baselinestretch}{0.928}
\newtheorem{theorem}{Theorem}

\newtheorem{lemma}{Lemma}
\newtheorem{definition}{Definition}
\makeatletter
\def\ifundefined{\@ifundefined}
\makeatother 

\begin{document}

\title{\vspace{-0.2cm}Laxity-Based Opportunistic Scheduling with Flow-Level Dynamics and Deadlines\vspace{-0.2cm}}

%\author{
%\authorblockN{}
%\authorblockA{} \and
%\authorblockN{}
%\authorblockA{} }
%
%\author{\IEEEauthorblockN{Huasen Wu\IEEEauthorrefmark{1},
%Xin Liu\IEEEauthorrefmark{2}\IEEEauthorrefmark{3}, and Youguang Zhang\IEEEauthorrefmark{1}}\\
%\IEEEauthorblockA{\IEEEauthorrefmark{1}School of Electronic and
%Information Engineering, Beihang University, Beijing 100191, China\\
%Email: huasenwu@gmail.com}\\
%\IEEEauthorblockA{\IEEEauthorrefmark{2}Microsoft Research Asia, Beijing 100080, China}\\
%\IEEEauthorblockA{\IEEEauthorrefmark{3}Department of Computer Science, University of California, Davis, CA 95616, USA}}

\author{
\IEEEauthorblockN{Huasen Wu and Youguang Zhang}
\IEEEauthorblockA{School of Electronic and Information Engineering\\
Beihang University, Beijing 100191, China\\
Email: huasenwu@gmail.com\vspace{-1cm}}
\and
\IEEEauthorblockN{Xin Liu}
\IEEEauthorblockA{Microsoft Research Asia, Beijing 100080, China\\
University of California, Davis, CA 95616, USA\vspace{-2cm}}}

\maketitle

\begin{abstract}
Many data applications in the next generation cellular networks, such as content precaching and video progressive downloading, require flow-level quality of service (QoS) guarantees. One such requirement is deadline, where the transmission task needs to be completed before the application-specific time. To minimize the number of uncompleted transmission tasks, we study laxity-based scheduling policies in this paper. We propose a Less-Laxity-Higher-Possible-Rate (L$^2$HPR) policy and prove its asymptotic optimality in underloaded identical-deadline systems. The asymptotic optimality of L$^2$HPR can be applied to estimate the schedulability of a system and provide insights on the design of scheduling policies for general systems. Based on it, we propose a framework and three heuristic policies for practical systems. Simulation results demonstrate the asymptotic optimality of L$^2$HPR and performance improvement of proposed policies over greedy policies.
\end{abstract}

%\begin{IEEEkeywords}
%
%\end{IEEEkeywords}

%%%%%%%%%%
\section{Introduction}\label{sec:introd}
%%%%%%%%%%
%Cellular operators are facing grand challenges in satisfying the exponentially increasing demand for wireless data service. Data-intensive mobile devices such as smartphones and tablets enable game-changing applications, such as video streaming and social networks, on mobile Internet  and impose great pressure on service providers \cite{Cisco2012WP}.

%However, {\it not all traffic is created equal}. For example, content precaching can tolerate minutes to hours of delay.
%%Usually, the users of such applications only care about the completion time of the entire transmission task but not the speed for transmitting each individual packet.
%Therefore, the delay-tolerant data can be leveraged to improve network resource utilization by opportunistically scheduling their transmission when the channel conditions are more favorable. Such precaching traffic is characterized by flow-level dynamics and deadlines. This is because scheduling such traffic needs to be carried out across a greater temporal scale, during which the population of users may change. On the other hand, the transmission tasks should be completed before their application-specific deadlines to maintain the required quality of experience (QoE). Policies for trading off between maximizing the system throughput and serving the most urgent users are required to minimize the probability of delay violation, and thus improve the network utilization and user experience.

Opportunistic scheduling plays an important role in improving network resource efficiency and user experience. A large number of scheduling policies, such as Proportional Fair (PF) scheduler \cite{Tse2001TR} and MaxWeight \cite{Andrews2004MLWDF}, have been proposed to balance between the system throughput and the level of satisfaction among different users. Most existing work focuses on the packet-level scheduling, where the number of users is assumed to be fixed and the performance is defined on the packet-level, e.g., number of packets received in a unit time or average delay of all received packets.

On the other hand, file download and multimedia streaming become increasing popular in cellular networks \cite{Evensen2010NOSSDAV}.
%The users of such applications do not care about the speed for transmitting each individual packet but require the transmission task to be completed before certain time. These data can be leveraged to improve network resource utilization by opportunistically scheduling their transmission when the channel conditions are more favorable.
The traffic generated by these applications is characterized by flow-level dynamics and deadlines. This is because scheduling such traffic needs to be carried out across a greater temporal scale, during which the population of users may change. In addition, the transmission tasks should be completed before their application-specific deadlines to maintain the required quality of experience (QoE). For example, in progressive downloading, to achieve quasi-live streaming, a segment of a video should be downloaded before the buffer depletes, which imposes a deadline of several seconds \cite{Evensen2010NOSSDAV}. Therefore, we study opportunistic scheduling policies to minimize the delay violation probability in wireless networks with flow-level dynamics and deadlines.

%Over the last decade, opportunistic scheduling has been extensively studied and standardized in all 3G/4G cellular networks. A large number of scheduling policies, such as Proportional Fair (PF) scheduler \cite{Tse2001TR} and MaxWeight \cite{Andrews2004MLWDF}, have been proposed to balance between the network resource efficiency and the level of satisfaction among different users. However, most existing work focuses on the packet-level scheduling, where the number of users is assumed to be fixed and the performance is defined on the packet-level, e.g., number of packets received in an unit time and average delay of all received packets. Hence, these policies cannot be directly applied in scheduling traffic with flow-level dynamics and deadlines.

Flow-level scheduling has been considered in the literature. Similar to the packet-level scheduling, a critical issue is to guarantee stability if possible. Recent results show that the maximum stable region can be easily achieved by applying some simple rules such as Best-Rate (BR) rule \cite{Ayesta2011ValueTools}. Other papers investigate policies for minimizing the average transmission delay. In \cite{Sadiq2010ITC:SRPTvsOPP,Aalto2011SIGMETRIX}, with the assumption of fast varying channel conditions, it is shown that combining opportunistic scheduling and the Shortest-Remaining-Processing-Time (SRPT) discipline in machine-job scheduling can minimize the average delay. However, the transmission delay may exceed the user's tolerant delay and become useless. In \cite{Proebster&Kaschub2011ICC}, the authors study flow-level scheduling policies for maximizing delay-dependent utility functions. This model can be viewed as scheduling with soft deadline constraints. However, it requires the knowledge of channel states in the future, which may be difficult in practice.

Scheduling with deadlines has been investigated in machine-job scheduling literature. Policies such as Earliest-Deadline-First (EDF) and Least-Laxity-First (LLF) have been proposed and shown to be optimal for underloaded systems \cite{Mok1983PhD}. Namely, a feasible schedule can be obtained by EDF and LLF if there are some off-line policies can do so. Other policies, e.g., D$^\text{over}$ \cite{Koren1995Dover}, have been proposed for overloaded systems and are shown to achieve the optimal competitive ratio. However, the temporal variation of data rate makes the design and analysis of scheduling policies for wireless networks with flow-level deadlines challenging. In \cite{Agarwal2002INFOCOM}, the authors show the Max C/I policy, which greedily serves the user with the highest data rate, achieves the optimal competitive ratio, assuming a partial value model. In this model, one user does not require the completion of the entire transmission task and the value is in proportion to the amount of data received. In many applications, however, it is required that at least certain percentage of data should be received or it will be useless.
%
%In \cite{Yang2011ICC:DPS}, a Dynamic Predictive Scheduling (DPS) is proposed for downlink scheduling in Drive-thru networks, where the vehicular users request downloading files in the coverage of an access point (AP) and the download task should be completed before they leave the system. However, it is assume that only the transmission data rate depends only on the distance between the vehicular user and the AP, and the fast fading effect is not taken into account.

In this paper, to minimize the number of uncompleted tasks, we study scheduling policies that balance between serving urgent users and maintaining multi-user diversity. We quantify the urgency of transmission tasks with laxity and propose laxity-based policies for scheduling file download traffic. Under the assumption of polymatroid capacity region \cite{Sadiq2010ITC:SRPTvsOPP}, we propose a Less-Laxity-Higher-Possible-Rate (L$^2$HPR) policy and show its asymptotic optimality in underloaded identical-deadline systems. To the best of our knowledge, this is the first theoretical  result on wireless scheduling with deadlines using the entire value model. The insights obtained from this policy can serve as a guideline to design policies for general systems. Based on it, we  propose a laxity-based policy framework and three heuristic policies for practical systems. Through numerical simulations, we demonstrate the asymptotic optimality of L$^2$HPR and the performance improvement of channel-and-urgency-aware policies.

%The remainder of the paper is organized as follows. In Section \ref{sec:syst_model}, we present the system model. In Section \ref{sec:asymp_opt_policy}, with the polymatroid capacity region assumption, we propose the L$^2$HPR policy and show its asymptotic optimality in underloaded identical-deadline systems. Inspired the design of L$^2$HPR and packet-level scheduling policies, we propose practical policies in Section \ref{sec:prac_heur_policy}. In Section \ref{sec:sim_res}, we carry out simulation studies comparing the performance of the proposed policies and some greedy policies. We discuss future work and conclude the paper in Section \ref{sec:conclusions}.

%%%%%%%%%%
\section{System Model}\label{sec:syst_model}
%%%%%%%%%%
We consider the flow-level scheduling with deadlines in the downlink of a single cell. A sequence of users enter the system and request to download files with deadlines. They depart upon task completion or delay violation. The objective of the base station (BS) is to minimize the number of uncompleted requests.

\subsection{Traffic and Channel Model}
Let $\mathcal{I}$ be the index set of all users entering the system.
%We consider both finite-user systems and stationary-arrival systems, with $\mathcal{I} = \{1, 2, \ldots, M\}$ for an $M$-user system and $\mathcal{I} = \mathbb{N}$ for a stationary-arrival system.
For each user $i \in \mathcal{I}$, the download request is represented by a triple $(A_i, F_{i,0},D_i)$, where $A_i$, $F_{i,0}$, and $D_i$ denote the arrival time, the initial file size (in bits), and the deadline, respectively. All $A_i$s, $F_{i,0}$s, and $D_i$s are random variables. The difference between the deadline and the arrival time, i.e., $D_i - A_i$, reflects the delay tolerance of user $i$. We focus on file download applications such as content precaching. Hence, we assume that the file size $F_{i, 0}$ is available as soon as user $i$ arrives.

%Each user arrives at a continuous random time and all the users are indexed in the order of their arrival times, i.e., for $i_1 < i_2$, $A_{i_1} \leq A_{i_2}$. For the $M$-user system, the arrival time $A_i$ are some random variables in an time interval. For the infinite-user system, we assume the users arrive according to a Poisson process with intensity $\lambda$, and thus $A_i$ is the $i$-th jumping time of the Poisson process.
%
%For the file size, we assume that $F_{i,0}$ ($i \in \mathcal{I}$) are independent random variables with density $f_{F_i}(\cdot)$. We focus on downloading files for delay tolerant applications such as content precaching. Hence, we assume that the file size can be calculated before downloading and the file size becomes known by the BS as long as user $i$ arrives. The deadline $D_i$ is also a random variable, with $D_i - A_i$ reflecting the delay tolerance of user $i$. The tolerated delay is related to both the file size and the data rate, and will be discussed later.

All data are transmitted over a wireless channel from the BS to each user using Time-Division Multiplexing (TDM). The channel condition for each user is time-varying and is modeled as a stationary stochastic process $R_i(t)$ ($t \in \mathbb{R}$), where $R_i(t) \geq 0$ denotes the instantaneous rate at which the BS can transmit to user $i$ at time $t$. We assume a wireless system with homogeneous channels, where $R_i(t)$s ($i \in \mathcal{I}$) are statistically identical with  $\bar{R}_i = \mathbf{E}[R_i(t)] = \bar{R}$ for all $i \in \mathcal{I}$. For a more practical system with heterogenous channels, we can transform the original system into an equivalent system with homogeneous channels using the scaling technique in \cite{Sadiq2010ITC:SRPTvsOPP}. Similar to \cite{Sadiq2010ITC:SRPTvsOPP}, we normalize the data rate and the file size with respect to the average rate $\bar{R}$.

%In order to characterize the delay tolerance of each user, we adopt the metric introduced in \cite{Bender1998SODA}, which is referred to as {\it stretch}. Specifically, the stretch of user $i$ is defined as the ratio between the practical delay and $\bar{D}'_{i,0}$, where $\bar{D}'_{i,0} = F_{i,0}/\bar{R}$ is the expected delay of user $i$ in an ideal situation where the entire channel is occupied by user $i$. Let $\xi_i > 1$ denote the maximum stretch accepted by user $i$. Then, the deadline $D_i$ is given by $D_i = A_i + \xi_i \bar{D}'_{i,0}$.

%Moreover, similar to many works on flow-level opportunistic scheduling \cite{}, we assume that one user requests at most one file for simplicity and the file sizes $F_{i,0}$ ($i = 1,2, \ldots$) are i.i.d. random variables with density $f_{F}(\cdot)$. For a user having multiple download requests, we can batch those files and treat them as a single request. If different requests have different deadlines, we can handle them in an iterative manner and treat different requests as different users.

\subsection{Scheduling Process}
 The BS schedules the transmission in a slotted manner. Time is divided into time slots of length $\Delta t$ and each slot is indexed by an integer $n$. In slot $n$, we let $\mathcal{Q}[n] \subseteq\mathcal{I}$ be the index set of users present in the system, and $Q[n] = |\mathcal{Q}[n]|$ be the number of users. For each user $i \in \mathcal{Q}[n]$, we denote its residual file size by $F_i[n]$.

 At the beginning of the $n$-th slot, i.e., $t_n = n \Delta t$, the BS allocates user $i$ with data rate $c_i[n]$. The rate vector, which consists of all $c_i[n]$s ($i \in \mathcal{Q}[n]$), stays in the capacity region corresponding to $\mathcal{Q}[n]$ \cite{Sadiq2010ITC:SRPTvsOPP}. Thus, the residual file size of user $i$ evolves as follows:
 \begin{equation} \label{eq:file_evolve}
F_i[n+1] = \max\{0, F_i[n] - c_i[n] \Delta t\},
\end{equation}
where the initial value of residual file size is $F_{i,0}$.

Note that the time scale separation \cite{Sadiq2010ITC:SRPTvsOPP} is applied here. In other words, we assume that the channel conditions $R_i(t)$s ($i \in \mathcal{I}$) vary infinitely fast. Then, a time slot can be divided into mini-slots, each of which is in the order of the channel coherence time. If the BS schedules in each mini-slot, then the data rates allocated to the present users in each slot can be averaged out and the rate vector stays in the capacity region. This assumption is critical because it captures the multi-user diversity effect in a tractable manner. However, we note that the time scale separation is highly ideal, especially when the slot length $\Delta t \to 0$. We will only use it for asymptotic analysis in Section \ref{sec:asymp_opt_policy}, and will relax this assumption when designing heuristic policies for practical systems in Section \ref{sec:prac_heur_policy}.

%We assume that all measurements of the network states are taken at the beginning of each time slot.

%The BS makes scheduling decisions at the beginning of each time slot and we assume that all measurements of the states are taken at the beginning of each time slot.

%We will follow the convention that using round bracket for continuous time variable and square bracket for discrete time variable, e.g., $R_i(t)$ is the data rate at time $t$ and $R_i[n]$ is the data rate in slot $t$.

%At the beginning of time slot $n$, the BS chooses which user to serve based on the network status, i.e., the present user index set $\mathcal{Q}[n]$, and $(F_i[n], D_i, R_i[n])$ for each user $i \in \mathcal{Q}[n]$. We assume that these user and channel states are perfectly known by the BS. This is reasonable since all the data are buffered in the BS and the channel conditions can be collected through feedback channels.

%Furthermore, let $V$ be the number of users missing their deadlines in an $M$-user system, and $\nu$ be the average number of users missing their deadlines in a unit time in the Poisson arrival system. The objectives of the BS can be stated by
%\begin{equation} \label{eq:sched_obj1}
%\min_{\Pi}~p_{viol} = \frac{\mathbf{E}[V]}{M}, ~\text{and}~\min_{\Pi}~p_{viol} = \frac{\nu}{\lambda}
%\end{equation}
%for the $M$-user system and the Poisson arrival system, respectively.

The objective of the BS is to minimize the number of users violating their deadlines. However, designing optimal policies for such a scheduling problem is challenging even with the time scale separation argument.  Therefore, in this paper, we first focus on the optimal policies for underloaded systems under certain additional assumptions. We call a policy {\it optimal in underloaded systems}, if a feasible schedule can be obtained by it whenever there are some off-line policies can do so, following the convention in machine-job scheduling \cite{Mok1983PhD,Koren1995Dover}. Then, we propose heuristic policies for more general systems and evaluate their performance through simulations.

%We summarize the notations used in this paper as follows.
%
%$t$: time;
%
%$\Delta t$: time slot length;
%
%$n$: time slot index;
%
%$i$: user index;
%
%$\mathcal{Q}(t)$ ($\mathcal{Q}[n]$): the index set of users present at time $t$ (in slot $n$);
%
%$A_i$: arrival time of user $i$;
%
%$D_i$: deadline of user $i$;
%
%$F_i(t)$ ($F_i[n]$): residual file size of user $i$ at time $t$ (in slot $n$);
%
%$R_i(t)$ ($R_i[n]$):  data rate of user $i$ at time $t$ (in slot $n$);

%%%%%%%%%%
\section{Asymptotically Optimal Policy in Identical-Deadline Systems}\label{sec:asymp_opt_policy}
%%%%%%%%%%
In this section, we study the opportunistic scheduling problem in an identical-deadline system with $M$ users, all of which request to download files before a same deadline, i.e., $D_i = D$ for $i \in \mathcal{I} = \{1, 2, \ldots, M\}$.
%We start with the case where all users arrive at the same time, with the assumption of $A_i = 0$ for all $i = 1, 2, \ldots, M$, and then extend the results to more general scenarios where users arrive at any time before the deadline $D$.

Due to the flow-level dynamics, designing optimal policies is  challenging even for the identical-deadline system. Motivated by the idea of polymatroid capacity region \cite{Sadiq2010ITC:SRPTvsOPP}, we propose a laxity-based policy, referred to as L$^2$HPR, and prove its asymptotic optimality.

\subsection{Polymatroid Capacity Region}
\begin{figure}[htbp]
\begin{center}
\includegraphics[angle = 0,width = 0.55\linewidth, height = 0.45\linewidth]{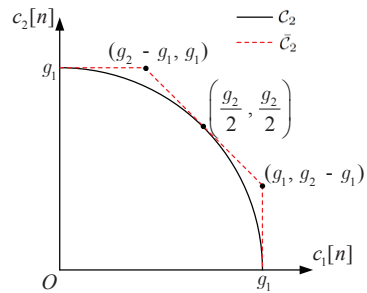}
\caption{Polymatroid capacity region for 2-user case \cite{Sadiq2010ITC:SRPTvsOPP}.} \label{fig:polymatroid_capacity_region}
\end{center}
\vspace{-0.8cm}
\end{figure}

To be self-contained, we briefly summarize the definition as follows. Polymatroid capacity region \cite{Sadiq2010ITC:SRPTvsOPP}  approximates the original capacity region with its polymatriod outer bound. Consider the scenarios where all the channel conditions are i.i.d. processes across all users. Let $g_k$ be the achievable multi-user diversity gain when there are $k$ active users, i.e., the ratio between the maximum throughput achieved by the $k$-user system and the single user system. Assume that $g_k$ is concavely increasing in $k$ and let $g_0 \equiv 0$. The polymatroid capacity region is defined as follows:
\begin{equation*} \label{eq:}
\bar{\mathcal{C}}_k \equiv \bigg\{\mathbf{r} \in [0,1]^k: \forall\mathcal{K} \subseteq\{1,\ldots,k\}, \sum_{i\in \mathcal{K}}r_i \leq g_{|\mathcal{K}|}\bigg\}. \nonumber
\end{equation*}
The polymatroid capacity region is the tightest polymatroid outer-bound region containing the original capacity region, which is shown in Fig.~\ref{fig:polymatroid_capacity_region} for the 2-user case. Thus, the minimum delay violation probability obtained with the polymatroid capacity region is a lower-bound of the practical system.

\subsection{Design of L$^2$HPR Policy}
%With the polymatroid capacity region, in time slot $n$, the BS chooses a rate vector $\mathbf{r}[n] \in \bar{\mathcal{C}}_{Q[n]}$ and serves the $j$-th ($j = 1, 2, \ldots, Q[n]$) user in $\mathcal{Q}[n]$, whose index is $i_{(j)}$, with data rate $\bar{R}_{i_{(j)}}[n] = r_j[n]$. Then, by modifying E.q. \eqref{eq:file_evolve}, we give the evolution of file size for user $i_{(j)}$ as follows
%\begin{equation} \label{eq:file_evolve_cont}
%F_{i_{(j)}}[n+1] = \max\{0, F_{i_{(j)}}[n] - r_j[n] \Delta t\},
%\end{equation}
%for $n \geq \lceil A_{i_{(j)}}/\Delta t \rceil$.

In order to minimize the number of uncompleted tasks, the BS should tradeoff between maintaining multi-user diversity (i.e., maximizing system throughput) and serving the more urgent users. To quantify the urgency of a given user, we introduce {\it expected laxity}.
The expected laxity is similar to the laxity defined in traditional job scheduling with the constant service rate replaced by the expected rate.
\begin{definition}(Expected Laxity) In slot $n$, for each $i \in \mathcal{Q}[n]$, the expected laxity is defined as
\begin{equation} \label{eq:exp_laxity}
L_i[n] = D_i - n \Delta t - \frac{F_i[n]}{g_1}.
\end{equation}
\end{definition}
In the above definition, the term $F_i[n]/g_1$ is the expected time required to finish the task with the entire channel allocated to user $i$. Hence, the expected laxity represents the time that can be allocated to other users without effecting the transmission of user $i$. Users with smaller expected laxity are more urgent.

Motivated by the intuition that the BS should allocate more resource to more urgent users, we propose a Less-Laxity-Higher-Possible-Rate (L$^2$HPR) policy.
\begin{definition}(L$^2$HPR policy) In slot $n$, when $\mathcal{Q}[n] \neq \emptyset$, sort all users in $\mathcal{Q}[n]$ in the ascending order of their expected laxity and let $j_{(i)}$ be the rank of user $i$. The L$^2$HPR policy serves each user $i \in \mathcal{Q}[n]$ with data rate
\begin{equation} \label{eq:}
c_i[n] = g_{j_{(i)}} - g_{j_{(i)} - 1}.
\end{equation}
\end{definition}

Note that with the assumption of concavely increasing gain, we have $g_{j} -  g_{j - 1} < g_{j+1} -  g_{j}$, indicating that with L$^2$HPR policy, the user with less expected laxity is allocated with higher data rate. Moreover, the total data rate is $\sum_{i \in \mathcal{Q}[n]} c_{i}[n] = g_{Q[n]}$, and thus the L$^2$HPR policy reaches the maximum system throughput that can be obtained when the number of users is $Q[n]$.

\subsection{Asymptotic Optimality of L$^2$HPR Policy}
In this subsection, we show that the L$^2$HPR policy is asymptotically (as the slot length $\Delta t$ tends to 0) optimal for underloaded systems with identical deadlines.

%We also remark that though the slot length $\Delta t$ tends to 0, we can still assume that $\Delta t$ is much larger than the time scale of channel variation. For example, for time $t > 0$, if $\Delta t = \frac{t}{N}$ and the channel variation scale is $\frac{t}{N^2}$, then, as $N$ tends to large, the slot length becomes large compared to the channel variation scale. Thus, the time scale separation argument is still valid when $\Delta t$ tends to 0.

It is worth noting that in constant-rate machine-job scheduling, the optimality of the LLF (or EDF) policy is shown with the exchange argument, i.e., transforming any feasible schedule into the one found by the LLF (or EDF) policy \cite{Mok1983PhD}. However, this approach does not work for L$^2$HPR since the feasible service rates depend on the number of users present in the system. Rather than using the exchanging argument, we study the asymptotic optimality by examining the least expected laxity obtained by L$^2$HPR in every time slot.

%In fact, the optimality of LLF can be viewed from another perspective: for any schedulable arrival sequence, the LLF policy obtains the maximum least-laxity in every slot and thus provides a feasible schedule (the proof can be found in our technical report \cite{}).

 We examine the state of all users having entered the system, including both present users and completed users. Let $\mathcal{Q}'[n]$ denote the index set of users arriving before time $t_n$, i.e.,
\begin{equation*} \label{eq:}
\mathcal{Q}'[n] = \{i: 1\leq i\leq M, A_i \leq t_n\}. \nonumber
\end{equation*}
We notice that by setting the residual size $F_i[n] = 0$, the expected laxity defined in \eqref{eq:exp_laxity} can also be applied for completed users. We further notice that the term $n\Delta t$ in \eqref{eq:exp_laxity} is common for all users. For the sake of notation simplicity, we introduce the {\it virtual expected laxity} for each user $i\in \mathcal{Q}'[n]$, which is defined as follows
\begin{equation*} \label{eq:}
L'_i[n] = L_i[n] + n\Delta t = D - \frac{F_i[n]}{g_1}. \nonumber
\end{equation*}
Note that by this definition, the virtual expected laxity is $L'_i[n] = D$ for completed users since $F_i[n] = 0$.

In order to show the asymptotic optimality of L$^2$HPR, we need to examine the least expected laxity, or equivalently, the least virtual expected laxity, obtained by L$^2$HPR in every slot. The {\it least virtual expected laxity} is given by
\begin{equation*} \label{eq:}
\underline{L}'_{\text{L$^2$HPR}}[n] = \min_{i \in \mathcal{Q}'[n]} L'_i[n],\nonumber
\end{equation*}
and the {\it least-laxity-user} is defined as the user having the least virtual expected laxity, i.e.,
\begin{equation*} \label{eq:}
i^*[n] = \argmin_{i \in \mathcal{Q}'[n]} L'_i[n].\nonumber
\end{equation*}
Note that there may be more than one user having the least virtual expected laxity and we let $i^*[n]$ be the one with the smallest index. This will not affect the analysis result since the performance of L$^2$HPR is reflected by the value of the least virtual expected laxity.

We expect that L$^2$HPR achieves the maximum least virtual expected laxity in every time slot to ensure the optimality of L$^2$HPR. Unfortunately, this is not always true because L$^2$HPR serves users with rate values from a discrete set $\{g_1, g_2-g_1, \ldots, g_M - g_{M-1}\}$, but other policies can reach larger least virtual expected laxity by using finer allocation. For example, for two users with $L'_1[0] = L'_2[0] = \frac{D}{2}$, the L$^2$HPR policy allocates data rates $g_1$ and $g_2 - g_1$ to these two users and the least virtual expected laxity of L$^2$HPR is $\underline{L}'_{\text{L$^2$HPR}}[1] = \frac{D}{2} + \frac{g_2-g_1}{g_1}\Delta t$. However, one can allocate equal data rate, i.e., $\frac{g_2}{2}$, to the two users and achieve the least virtual expected laxity $\frac{D}{2}+\frac{g_2}{2 g_1}\Delta t$. This is larger than $\underline{L}'_{\text{L$^2$HPR}}[1]$ due to the concavely increasing property of $g_k$. Nevertheless, we can show that the difference vanishes as the slot length $\Delta t$ tends to 0 and thus L$^2$HPR is asymptotically optimal.

First, we focus on the case where all users arrive at the same time. Without loss of generality, we assume $A_i = 0$ for all $i = 1,2,\ldots, M$. Hence, for all $n \geq 0$, $\mathcal{Q}'[n] = \{1,2,\ldots,M\}$.

Recall that the key idea of the L$^2$HPR policy is allocating larger data rate to the user with less virtual expected laxity. Thus, we first define two relationships, ``Used-to-be-Less-Than (ULT)'' and ``Indirectly-Used-to-be-Less-Than (I-ULT)'', which will play an important role in analyzing the performance of L$^2$HPR.
\begin{definition}(ULT and I-ULT)

a) In slot $n$, for two users $i_1$ and $i_2$, we say that $i_1$ used-to-be-less-than (ULT) $i_2$ before slot $n$, denoted as $i_1 \leq_n i_2$, if there exists an $n'$ ($0\leq n'\leq n$) such that $L'_{i_1}[n'] \leq L'_{i_2}[n']$.

b) In slot $n$, for two users $i_1$ and $i_2$, we say that $i_1$ indirectly-used-to-be-less-than (I-ULT) $i_2$ before slot $n$, denoted as $i_1 \preceq_n i_2$, if there exists a user sequence $i'_1,i'_2, \ldots, i'_m$, such that $i_1 \leq_n i'_1$, $i'_1 \leq_n  i'_2$, $\ldots$ , and $i'_m \leq_n i_2$.
\end{definition}

%Note that the relationship ULT is not transitive. Namely, for three different users, $i_1$, $i_2$, and $i_3$, the conditions $i_1 \leq_n i_2$ and $i_2 \leq_n i_3$ does not necessary imply $i_1 \leq i_3$, since the events about their virtual expected laxity, $L'_{i_1} \leq L'_{i_2}$ and $L'_{i_2} \leq L'_{i_3}$ may occur in different time slots. However, according to the definition of I-ULT, we can easily see that I-ULT is a transitive relationship.

The lemma below shows that under the L$^2$HPR policy, if a user ULT another user, its virtual expected laxity will not much exceed that of the other user.
\begin{lemma}\label{thm:laxity_diff}
For an identical-deadline system under the L$^2$HPR policy, if users $i_1$ and $i_2$ satisfy $i_1 \leq_n i_2$, then
\begin{equation} \label{eq:laxity_diff}
L'_{i_1}[n] - L'_{i_2}[n] \leq \Delta t.
\end{equation}
\end{lemma}
Lemma \ref{thm:laxity_diff} can be proved by using the definition of ULT and tracing the virtual expected laxity of users $i_1$ and $i_2$. It is omitted here due to the space limitation. Interested users are referred to Appendix \ref{app:proof_of_laxity_diff}.

%Under the L$^2$HPR policy, the least-laxity-user may change across time slots.
In order to provide a lower bound on the least virtual expected laxity obtained by L$^2$HPR, we define a {\it least-laxity-set}, which contains the least-laxity-user $i^*[n]$ and all other users I-ULT $i^*[n]$.
\begin{definition}(Least-Laxity-Set) The least-laxity-set in slot $n$ is an index set $\underline{\mathcal{Q}}[n]$ satisfying the following conditions:

a) $i^*[n] \in \underline{\mathcal{Q}}[n]$;

b) For any $i \in \mathcal{Q}[n]$, $i \in \underline{\mathcal{Q}}[n]$ if and only if $i \preceq_n i^*[n]$.
\end{definition}

Let $\underline{Q}[n] = |\underline{\mathcal{Q}}[n]|$ be the number of elements in $\underline{\mathcal{Q}}[n]$. By the definition of  $\underline{\mathcal{Q}}[n]$, we know that from time slot $0$ to time slot $n-1$, all the $\underline{Q}[n]$ largest data rates, i.e., $\{g_1, g_2 - g_1,\ldots,  g_{\underline{Q}[n]} - g_{\underline{Q}[n]-1}\}$, are allocated to users in  $\underline{\mathcal{Q}}[n]$. With this property of $\underline{\mathcal{Q}}[n]$, we present the following lemma stating a lower bound on the least virtual expected laxity of L$^2$HPR.

\begin{lemma}\label{thm:laxity_bound}
The least virtual expected laxity obtained by L$^2$HPR in time slot $n$ is bounded as
\begin{eqnarray} \label{eq:laxity_bound_slotted}
\underline{L}'_{\text{L$^2$HPR}}[n] \geq \min\bigg\{D - (M-1)\Delta t, ~~~~~~~~~~~~~~~~~~~~\nonumber\\
\frac{1}{\underline{Q}[n]}\bigg[\sum_{i\in \underline{\mathcal{Q}}[n]}L'_i[0] + \frac{n g_{\underline{Q}[n]}\Delta t}{g_1}\bigg] - (M-1)\Delta t\bigg\}.
\end{eqnarray}
\end{lemma}
\begin{proof}
By the definition of I-ULT, for any $i \in \underline{\mathcal{Q}}[n]$, $i \neq i^*[n]$, we can find a sequence of different users, $\{i'_1, i'_2, \ldots, i'_m\}$, such that $i \leq_n i'_1$, $i'_1 \leq_n  i'_2$, $\ldots$ , and $i'_m \leq_n i^*[n]$. Note that the sequence does not contain $i$ or $i^*[n]$ and thus $m < M-2$. According Lemma \ref{thm:laxity_diff}, we have
\begin{equation} \label{eq:bound_laxity_diff}
L'_i[n] \leq \underline{L}'_{\text{L$^2$HPR}}[n] +  (M-1)\Delta t.
\end{equation}

Consequently, when there are some completed users in $\underline{\mathcal{Q}}[n]$, the least virtual expected laxity is bounded by
\begin{equation} \label{eq:least_laxity_somecompleted}
\underline{L}'_{\text{L$^2$HPR}}[n] \geq D - (M-1)\Delta t.
\end{equation}

On the other hand, when there are no completed users in $\underline{\mathcal{Q}}[n]$, as we have pointed out before, from time slot $0$ to $n-1$, all the $\underline{Q}[n]$ largest data rates are allocated to the users in  $\underline{\mathcal{Q}}[n]$. Thus, the sum of virtual expected laxity is
\begin{equation} \label{eq:sum_laxity_noncompleted}
\sum_{i\in \underline{\mathcal{Q}}[n]} L'_i[n] = \sum_{i\in \underline{\mathcal{Q}}[n]} L'_i[0] + \frac{n g_{\underline{Q}[n]}\Delta t}{g_1}.
\end{equation}
Note that from \eqref{eq:bound_laxity_diff}, we have
\begin{equation}\label{eq:least_laxity_noncompleted}
\underline{L}'_{\text{L$^2$HPR}}[n] \geq \frac{1}{\underline{Q}[n]} \sum_{i\in \underline{\mathcal{Q}}[n]} L'_i[n] - (M-1)\Delta t.
\end{equation}

Finally, combining \eqref{eq:least_laxity_somecompleted}, \eqref{eq:sum_laxity_noncompleted}, and \eqref{eq:least_laxity_noncompleted} , we know that \eqref{eq:laxity_bound_slotted} is true.
\end{proof}

Next, using Lemma \ref{thm:laxity_bound}, we show the asymptotic (as the slot length $\Delta t \to 0$) optimality of L$^2$HPR in underloaded identical-deadline systems.
\begin{theorem} \label{thm:asymp_opt_LLHPR_identical_arr}
Assume that for $i = 1, 2, \ldots, M$, $A_i = 0$, $D_i = D >0$. When the slot length $\Delta t \to 0$, the L$^2$HPR policy achieves the maximum least-laxity at any time $t$ and is asymptotically optimal in underloaded systems.
\end{theorem}
\begin{proof} For given $t$, we divide the time into $N$ slots and the slot length $\Delta t = t/N$. According to Lemma \ref{thm:laxity_bound}, as $N \to \infty$ and $\Delta t \to 0$, the least virtual expected laxity at time $t$ satisfies
\begin{equation} \label{eq:laxity_bound_cont}
\underline{L}'_{\text{L$^2$HPR}}[N] \geq \min\bigg\{D, \frac{1}{\underline{Q}[N]}\bigg[\sum_{i\in \underline{\mathcal{Q}}[N]}L'_i[0] + \frac{g_{\underline{Q}[N]}t}{g_1}\bigg]\bigg\}.
\end{equation}
Because $g_{\underline{Q}[N]}t$ is the maximum throughput that $\underline{Q}[N]$ users can obtain in a duration $t$, there are no other feasible schedules can obtain larger least virtual expected laxity than L$^2$HPR. Consequently, when the arrival sequence is schedulable, the L$^2$HPR policy will generate a feasible schedule as $\Delta t \to 0$ and is asymptotically optimal in underloaded systems.
\end{proof}

This conclusion can be extended to the case with identical deadlines but different arrival times, which is stated by the following theorem.
\begin{theorem} \label{thm:asymp_opt_LLHPR_diff_arr}
Assume that for $i = 1, 2, \ldots, M$, $D_i = D$, $A_i < D$. As the slot length $\Delta t \to 0$, the L$^2$HPR policy achieves the maximum least-laxity at any time $t$ and is asymptotically optimal in underloaded systems.
\end{theorem}
The proof of Theorem \ref{thm:asymp_opt_LLHPR_diff_arr} is similar to Theorem \ref{thm:asymp_opt_LLHPR_identical_arr}. The main difference is that in the different-arrival-time case, since there will be some new arrivals, the least-laxity-set $\underline{\mathcal{Q}}[n]$ is not monotonically increasing, i.e., $\underline{\mathcal{Q}}[n] \subseteq \underline{\mathcal{Q}}[n+1]$ is not necessarily true. Thus, we need to discuss the laxity in different temporal intervals.  Interested readers are referred to Appendix \ref{app:proof_of_asymp_opt_LLHPR_diff_arr}.

%%%%%%%%%%
\section{Practical Heuristic Laxity-based Policies}\label{sec:prac_heur_policy}
%%%%%%%%%%
The asymptotically optimal policy L$^2$HPR is based on the idealized assumption of polymatroid capacity region, and cannot be implemented in practical systems. In this section, we propose practical heuristic laxity-based policies.

\subsection{Policy Structure}
First, for a TDM system, typically at most one user can be served in each time slot. We assume that the slot length $\Delta t$ is sufficiently small and the channel state is constant within one time slot. Let $R_i[n] = R_i(t_n)$ be the data rate supported by user $i$ in slot $n$. Then, in slot $n$, using policy $\Pi$, the BS chooses user $\Pi(S[n])$ to serve based on the network status $S[n]$, which is given by
\begin{equation*} \label{eq:}
S[n] = \{\mathcal{Q}[n]; (A_i, F_i[n], D_i), i \in \mathcal{Q}[n]; R_i[n], i \in \mathcal{Q}[n]\}.   \nonumber
\end{equation*}

%From the design and analysis of L$^2$HPR, we know that in order to minimize the delay violation probability, the BS needs to tradeoff between maximizing the system throughput and maximizing the least expected laxity. For the former part, the BS would like to serve the user with the highest instantaneous data rate, while for the latter part, the BS prefers to serve the most urgent user, which have the least expected laxity.
%%Inspired by the policies for packet level scheduling, which make the scheduling decision according to some weights, we define the {\it weight} in our system as
%%\begin{equation} \label{eq:weight_general}
%%W_i[n] = \kappa_i R_i[n] U(L_i[n]),~~ i = 1, 2, \ldots, Q[t],
%%\end{equation}

Furthermore, another issue is that a practical system may be overloaded, i.e., not all download tasks can be completed before their deadlines. Serving users which are likely to expire may waste the chance to finish other download tasks. Thus, according to the expected laxity, we divide the present users $\mathcal{Q}[n]$ into two groups, $\mathcal{G}_{\delta}^{(+)}[n]$ and $\mathcal{G}_{\delta}^{(-)}[n]$, which are given by
\begin{equation*} \label{eq:}
\mathcal{G}_{\delta}^{(+)}[n] = \{i: i\in \mathcal{Q}[n], L_i[n] \geq \delta\}, \nonumber
\end{equation*}
and
\begin{equation*} \label{eq:}
\mathcal{G}_{\delta}^{(-)}[n] = \{i: i\in \mathcal{Q}[n], L_i[n] < \delta\}. \nonumber
\end{equation*}
The users in $\mathcal{G}_{\delta}^{(+)}[n]$ will be served by trading off between the data rate and the urgency. For the users in $\mathcal{G}_{\delta}^{(-)}[n]$, they will be served only if $\mathcal{G}_{\delta}^{(+)}[n] = \emptyset$, so that we do not waste on tasks that are unlikely to be finished. Moreover, for $\mathcal{G}_{\delta}^{(-)}[n]$, we will simply serve the user with the highest data rate to maximize the system throughput. Specifically, we propose the following policy framework:
\begin{eqnarray} \label{eq:policy_struct}
\Pi(S[n])=
\begin{cases}
\underset{i \in \mathcal{G}_{\delta}^{(+)}[n]}{\argmax}~\kappa_i R_i[n] U(L_i[n]),&\text{if $\mathcal{G}_{\delta}^{(+)}[n] \neq \emptyset$},\\
\underset{i \in \mathcal{G}_{\delta}^{(-)}[n]}{\argmax}~\kappa_i R_i[n],&\text{if $\mathcal{G}_{\delta}^{(+)}[n] = \emptyset$},
\end{cases}
\end{eqnarray}
where $\kappa_i >0$ is a constant for distinguishing priorities of different applications, and $U(\cdot)$ is a decreasing function that quantifies the urgency based on the expected laxity $L_i[n]$ defined by \eqref{eq:exp_laxity}.
With the structure proposed above, we can obtain different policies by designing different urgency functions.

\subsection{Laxity-based Heuristic Policies}
We construct the urgency function in \eqref{eq:policy_struct} to obtain different policies. First, note that when $\delta < 0$, the expected laxity $L_i[n]$ for $i \in \mathcal{G}_{\delta}^{(+)}[n]$ may be negative. We deal with this issue by using an approximation $L_i^\epsilon[n]$, which is given by
\begin{equation*} \label{eq:}
L_i^\epsilon[n] = \max\{L_i[n], \epsilon\}, \nonumber
\end{equation*}
where $\epsilon > 0$ is a small constant.

We propose three heuristic polices based on polynomial, exponential, and logarithm urgency function,  and refer  to them as L-MaxWeight, L-Exp, and L-Log, respectively.

{\it a) L-MaxWeight}
\begin{equation*} \label{eq:}
U_{\text{L-MaxWeight}}(L_i[n]) = (L_i^\epsilon[n])^{-\alpha}, \nonumber
\end{equation*}
where $\alpha > 0$.

{\it b) L-Exp}
\begin{equation*} \label{eq:}
U_{\text{L-Exp}}(L_i[n]) = \exp\left\{-\frac{\beta_i L_i^\epsilon[n]}{\zeta + (\bar{L}[n])^\eta}\right\}, \nonumber
\end{equation*}
where $\beta_i > 0$, $\zeta > 0$, and $\bar{L}[n] = \frac{1}{|\mathcal{G}_{\delta}^{(+)}[n]|}\sum_{i \in \mathcal{G}_{\delta}^{(+)}[n]}\beta_i L_i^\epsilon[n]$.

{\it c) L-Log}
\begin{equation*} \label{eq:}
U_{\text{L-Log}}(L_i[n]) = \frac{1}{\log(\zeta + \beta_i L_i^\epsilon[n])}, \nonumber
\end{equation*}
where $\beta_i > 0$ and $\zeta > 0$.

Rigorous analysis for the performance of the above heuristic policies is challenging and we will evaluate their performance through simulations in the next section.

\section{Simulation Results}\label{sec:sim_res}
%%%%%%%%%%
In this section, we evaluate the performance of the proposed laxity-based policies through simulations. We present simulation results on the schedulability and delay violation probability.
%
%We compare the performance of L$^2$HPR, L-MaxWeight, L-Exp, and L-Log. In addition, we also run simulations for the Max C/I policy \cite{Agarwal2002INFOCOM} which focuses on maximizing the system throughput , and EDF and LLF \cite{Mok1983PhD} which focus on serving the most urgent users.
\subsection{Simulation Settings}
\subsubsection{Traffic Model}
%We consider both the identical-deadline system and the stationary-arrival system.
We consider both identical-deadline systems with finite users and stationary-arrival systems. In the identical-deadline system with deadline $D$, we assume that there are $M$ users with arrival times uniformly distributed in the interval $[0, aD]$. $a \in [0, 1)$ is a constant for controlling the distribution of the arrival time.

In the stationary-arrival case, we assume that the users arrive according to a Poisson process with rate $\lambda$. Moreover, for each $i \in \mathcal{I}$, we set the deadline $D_i = A_i + \xi F_{i,0}/g_1$, where $\xi > 1$ is the maximum acceptable stretch factor. Note that the stretch factor is defined as the ratio between the practical delay and the expected delay in an ideal situation where the entire channel is occupied by a given user \cite{Bender1998SODA}. Hence, the metric $\xi$ provides an indication of how much delay the application can tolerate.

For the file size, we apply the model proposed in \cite{NGMN2008WP} for FTP traffic, where the file size follows truncated lognormal distribution with mean 2 Mbytes, standard deviation 0.722 Mbytes, and maximum size 5 Mbytes. The size is normalized with respect to the average rate.

%With regards to the file size, we apply the model proposed in \cite{NGMN2008WP} for FTP traffic, where the file size follows truncated lognormal distribution with mean 2 Mbytes, standard deviation 0.722 Mbytes, and maximum size 5 Mbytes.
%Namely, the probability density function is \cite{\cite{NGMN2008WP}}
%\begin{equation} \label{eq:}
%f_F(x) = \frac{1}{\sqrt{2\pi}\sigma x}\exp\big[-\frac{(\ln x - \mu)^2}{2 \sigma^2}\big], \nonumber
%\end{equation}
%where the size $x$ is measured in bytes, $0 < x \leq 5.0 \times 10^6$, $\sigma = 0.35$, and $\mu = 14.45$.

\subsubsection{Channel Model}
We use a continuous transmission rate model \cite{Sadiq2011ToN:LogRule,Proebster&Kaschub2011ICC} and assume that the data rate supported by each user is given by $R_i[n] = B\log_2\big(1 + \gamma_i[n]\big)$,
where $B$ is the bandwidth, and $\gamma_i[n]$ is the received SINR of user $i$ in slot $n$. Rayleigh fading channel is assumed for each user and thus the received SINR $\gamma_i[n]$  follows exponential distribution. We set $B = 800$ KHz and the expectation of SINR $\mathbf{E}\gamma_i[n] = 0$ dB. Note that these values only have a marginal effect since the data rate is normalized with respective to the average rate.

To obtain multi-user diversity gains used in L$^2$HPR, we set $g_1 = 1$ due to the normalization of the data rate, and obtain $g_k$ ($k > 1$) by evaluating the throughput of the Max C/I scheduler with $k$ users.

\subsubsection{Policy Parameters}

The parameters used by the proposed laxity-based policies are summarized in Table \ref{tab:plocy_param}. We set $\delta = -2 < 0$ since the users with negative (but not too large absolute value) expected laxity may still be able to be completed if they experience good channel conditions in the following slots. Other parameters are similar to those in conventional packet level scheduling policies \cite{Sadiq2011ToN:LogRule}.
%We note that the performance of the policies is robust with respect to moderate changes of these parameters.

\vspace*{-5pt}
\begin{table}[!h]
\tabcolsep 0pt \caption{Parameters used by scheduling policies}
\vspace*{-20pt}
\begin{center}
\def\temptablewidth{0.45\textwidth}
{\rule{\temptablewidth}{0.5pt}}
\begin{tabular*}{\temptablewidth}{@{\extracolsep{\fill}}p{2cm}p{2cm}p{2cm}p{2cm}}
Common &L-MaxWeight & L-Exp & L-Log \\
\hline
$\kappa_i = \frac{1}{\bar{R}_i}$ & $\alpha = 1$ & $\beta_i = 0.05$ & $\beta_i = 10$\\
$\delta = -2$&  &$\zeta = 1$ & $\zeta = 10$\\
$\epsilon = 0.001$& &$\eta = 0.5 $ &
\end{tabular*}
{\rule{\temptablewidth}{0.1pt}}
\end{center}
\vspace*{-28pt}
\label{tab:plocy_param}
\end{table}

\subsection{Schedulability under Different Policies}

\begin{figure}[thbp]
\begin{center}
\subfigure[Identical-deadline system]{
\includegraphics[angle = 0,width = 0.465\linewidth, height = 0.45\linewidth]{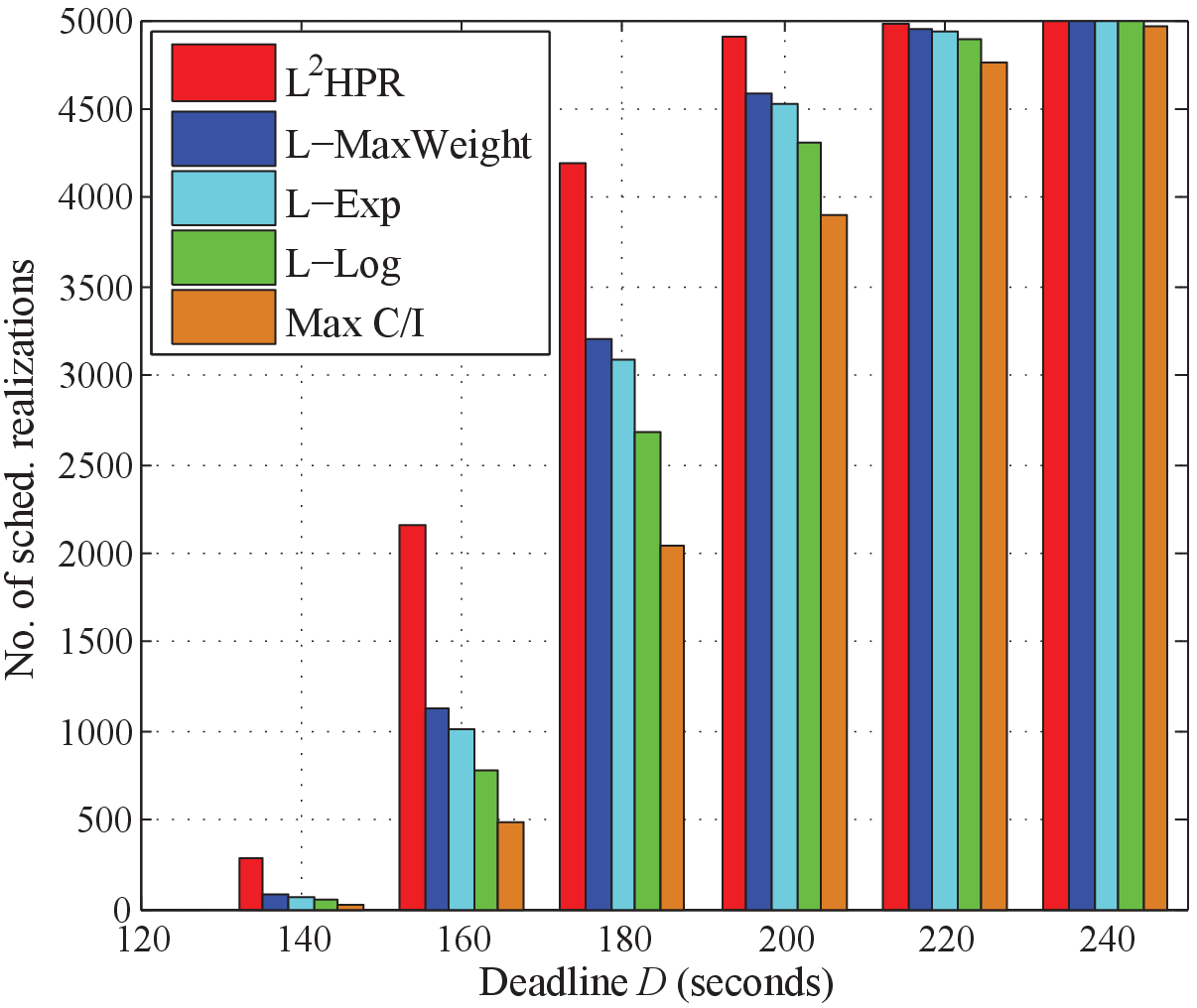}
\label{fig:identical_deadline_sched_num} }
\subfigure[Stationay arrival system]{
\includegraphics[angle = 0,width = 0.465\linewidth, height = 0.45\linewidth]{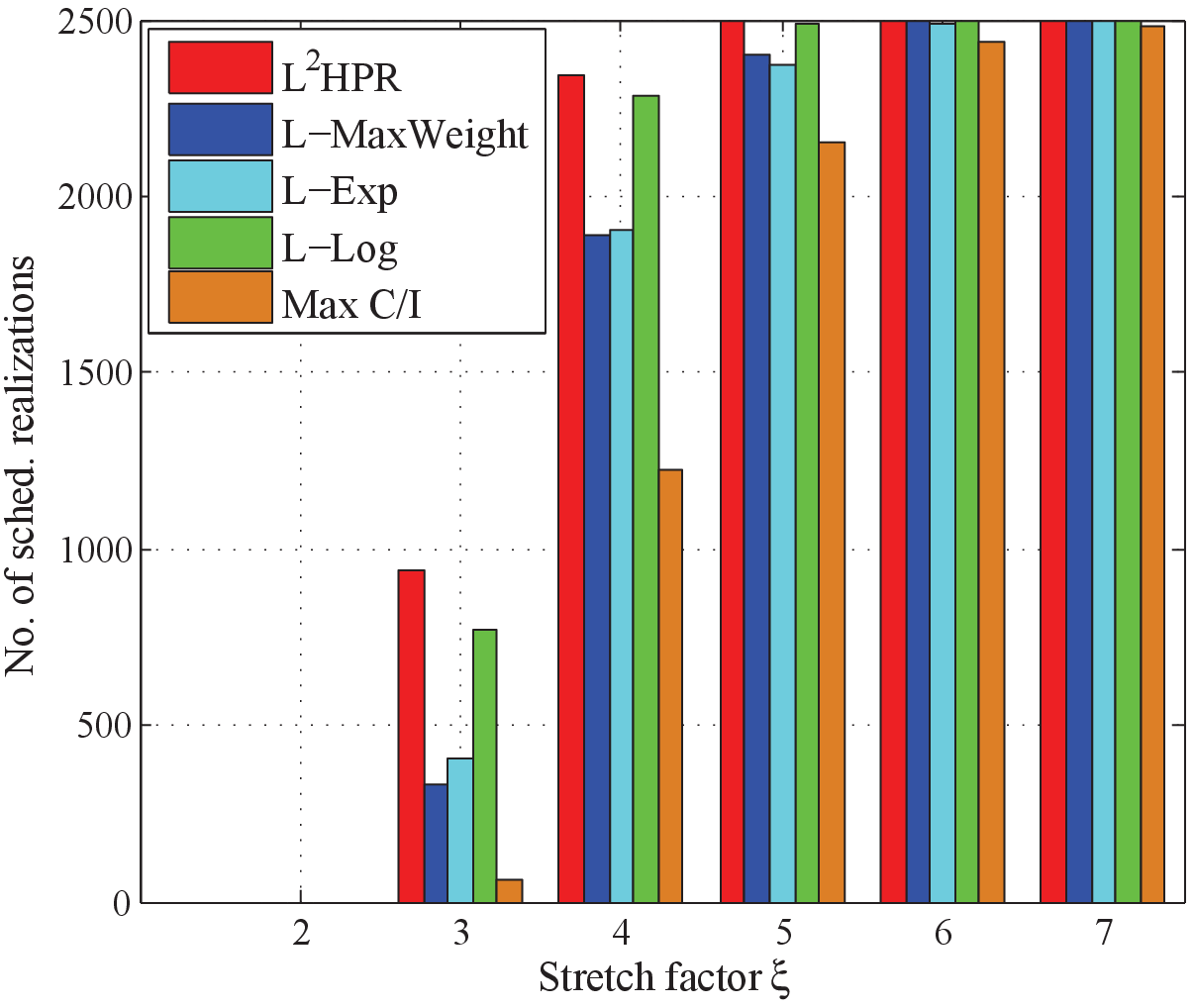}
\label{fig:dynamic_sched_num}}
\caption{Number of schedulable realizations ($M = 15$, $a = 0.5$, and $\lambda = 0.05$).}
\label{fig:sched_num}
\end{center}
\vspace{-0.8cm}
\end{figure}

Fig. \ref{fig:sched_num} shows the number of schedulable realizations under different scheduling policies.
%A realization is called schedulable under certain policy if all transmission tasks are completed before their deadlines.
From Figs. \ref{fig:identical_deadline_sched_num} and \ref{fig:dynamic_sched_num}, we can see that the proposed L$^2$HPR achieves the largest number of schedulable realizations among all policies. By tracing the scheduling results of each realization, we find that in the identical deadline system, a realization is schedulable under L$^2$HPR as long as it is schedulable under some other policies. In contrast, a schedulable realization under L$^2$HPR is not necessarily schedulable under other policies. These results demonstrate the asymptotic optimality of L$^2$HPR in underloaded identical-deadline system. Similar phenomenon occurs in the stationary-arrival system, though we cannot prove the asymptotic optimality of L$^2$HPR in such a general system.

The proposed laxity-based policies, L-MaxWeight, L-Exp, and L-LLF, outperform the greedy policy Max C/I in maximizing the number of schedulable realizations. Comparing the performance of L-MaxWeight, L-Exp, and L-LLF in the identical-deadline system and the stationary-arrival system, we can see that L-MaxWeight and L-Exp outperform L-LLF in the identical-deadline system, but the situation is reversed in the stationary-arrival system. This is because the variance of the expected laxity in the identical-deadline system is much smaller than that in the stationary-arrival system. The urgencies quantified by the logarithm function are similar among different users and L-LLF behaves as Max C/I in the identical-deadline system. But the L-LLF policy provides a better tradeoff when the variance of the expected laxity is large and performs better in the stationary-arrival system.

We emphasize that in the presented range, no realization is schedulable under EDF and LLF, which are unaware of the channel conditions (EDF is not evaluated in the identical-deadline system since all users have the same deadline). Even Max C/I, which makes scheduling decisions based only on channel conditions, can perform much better than EDF and LLF. This shows the value of channel condition knowledge in improving the scheduling performance.

\subsection{Delay Violation Probability}

\begin{figure}[thbp]
\begin{center}
\subfigure[Identical-deadline system]{
\includegraphics[angle = 0,width = 0.465\linewidth, height = 0.45\linewidth]{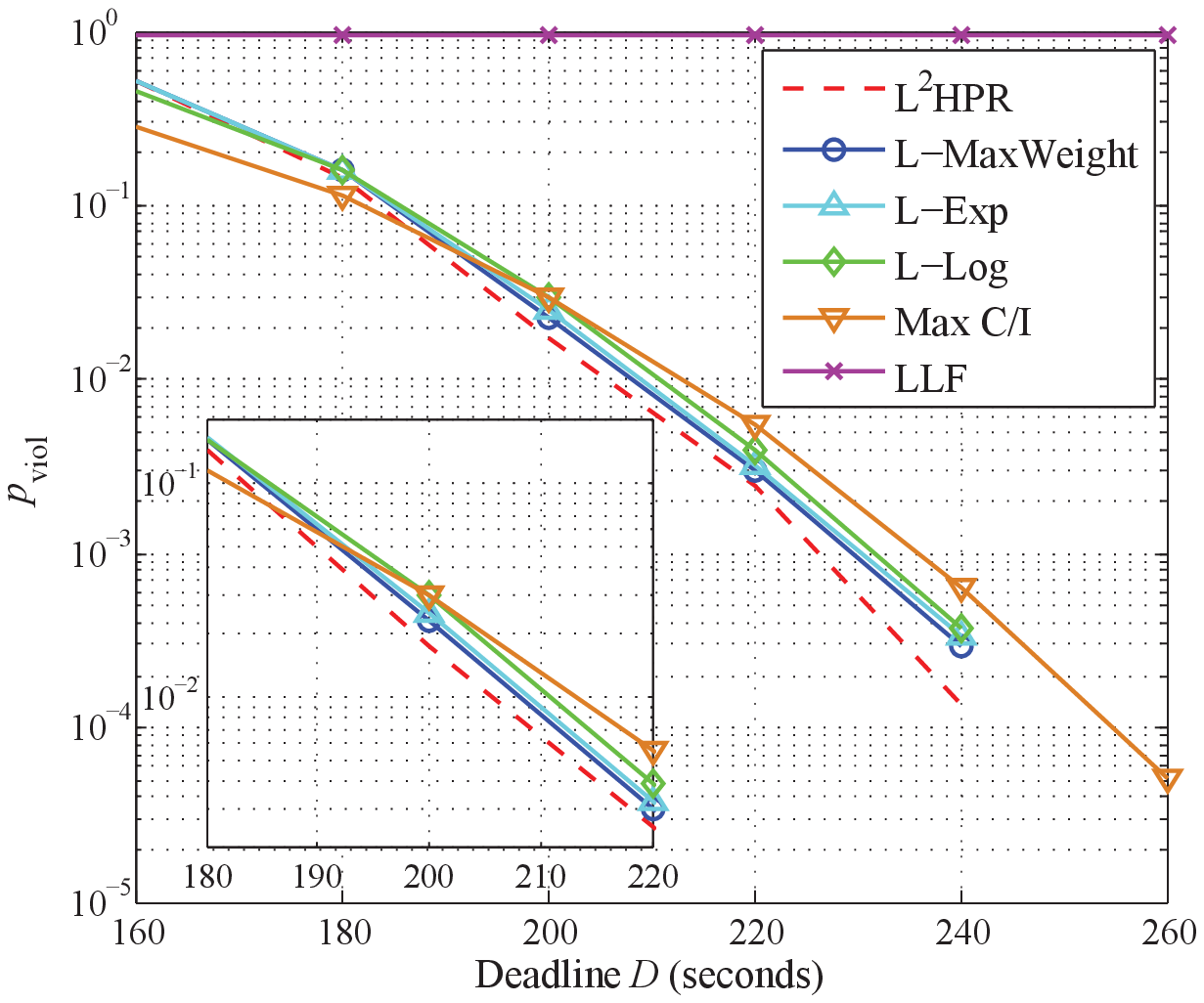}
\label{fig:identical_deadline_viol_prob} }
\subfigure[Stationary-arrival system]{
\includegraphics[angle = 0,width = 0.465\linewidth, height = 0.45\linewidth]{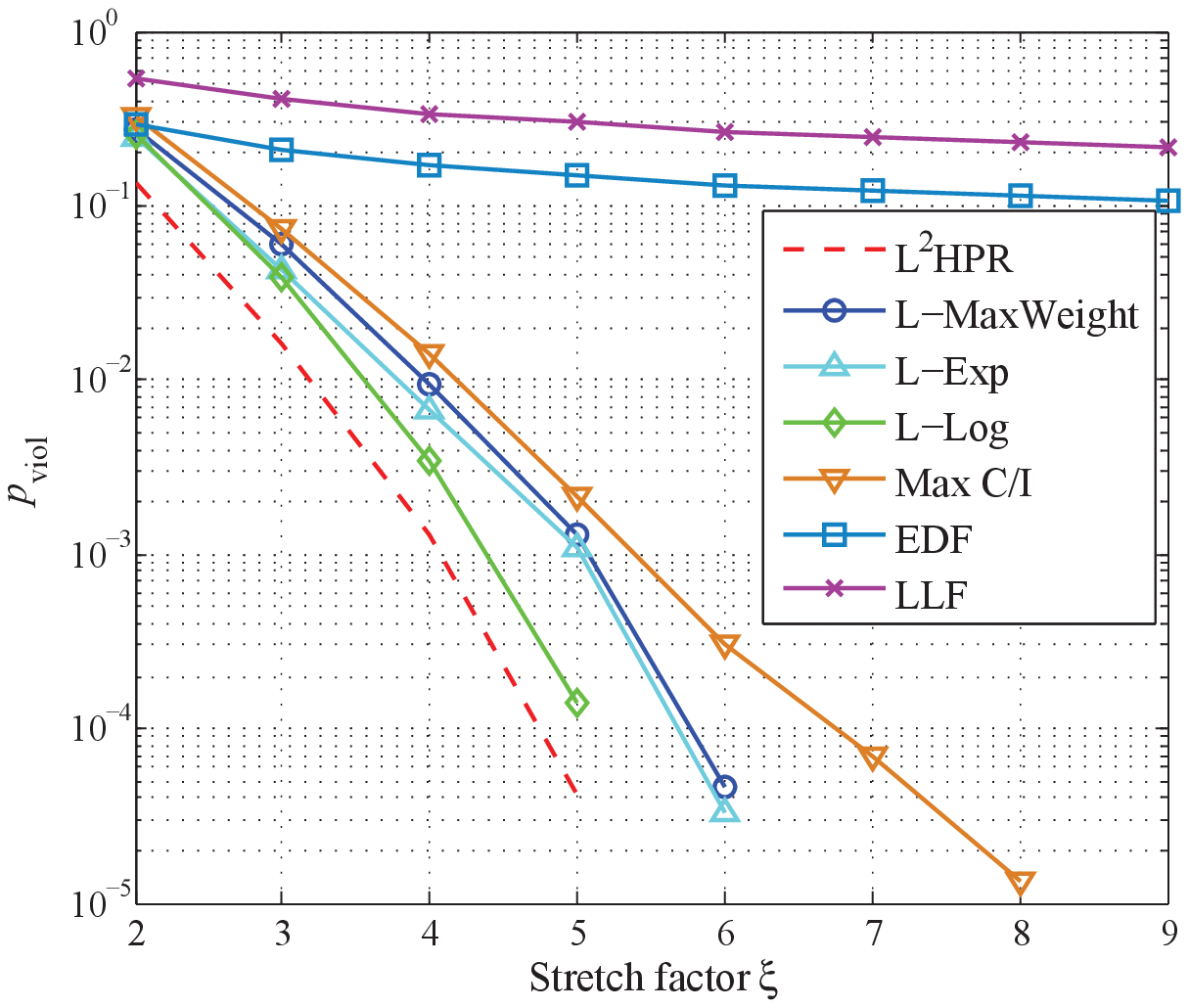}
\label{fig:dynamic_viol_prob}}
\caption{Delay violation probability ($M = 15$, $a = 0.5$, and $\lambda = 0.05$).}
\label{fig:viol_prob}
\end{center}
\vspace{-0.8cm}
\end{figure}

Fig. \ref{fig:viol_prob} depicts the delay violation probability under different policies. It is surprising at the first sight that from Fig. \ref{fig:identical_deadline_viol_prob}, the delay violation probability of L$^2$HPR is larger than that of Max C/I  when $D < 200$. This is because the L$^2$HPR policy tries to maximize the least expected laxity of the system by prioritizing the most urgent user. Thus, when a realization is unschedulable, resource is wasted and many users will violate their deadline constraints. Similar problems occur in other laxity-based policies.

In the stationary-arrival system, the proposed laxity-based policies outperform the greedy Max C/I policy. For example, the delay violation probability of L-LLF is only 25\% of that of Max C/I. In addition, the delay violation probability turns to be about $10^{-4}$ under L$^2$HPR and L-LLF when $\xi = 5$. While similar probability is obtained by Max C/I until $\xi = 7$, which requires additional 40\% delay.

Again, we point out that the channel-oblivious policies, EDF and LLF, perform rather badly compared with the channel-aware policies. Particularly, in the identical-deadline system under LLF, many users achieve very close expected laxity. Hence, most of users miss their deadlines and the delay violation probability only slightly decreases as $D$ increases.

%%%%%%%%%%
\section{Conclusion and Future Work}\label{sec:conclusions}
%%%%%%%%%%
In this paper, we study laxity-based policies for scheduling file downloading traffic which is characterized with flow-level dynamics and deadlines. Under the idealized assumption of polymatroid capacity region, we propose an asymptotically optimal policy, referred to as L$^2$HPR. We also propose heuristic policies, L-MaxWeight, L-Exp, and L-LLF, for practical systems. Comparative study between the proposed laxity-based policies and traditional greedy policies such as Max C/I, EDF, and LLF, demonstrates that the performance can be improved by intelligently balancing the multi-user diversity and the urgent users.

We mainly focus on designing optimal policies for underloaded systems and assume that all completed tasks have same value. However, in practice, it is possible that not all tasks can be finished before their deadlines and different tasks may have different values. In the future, we will study algorithms that estimate the schedulability of the coming sequence, drop users to avoid resource wasting, and maximize obtained utility when the system is possibly overloaded.

%%%%%%%%%%%
\appendices
%%%%%%%%%%%
%%%%%%%%%%%
\section{Proof of Lemma \ref{thm:laxity_diff}}\label{app:proof_of_laxity_diff}
%%%%%%%%%%%
We can show that for an identical-deadline system under L$^2$HPR, in any time slot $n'$, if $L'_{i_1}[n'] - L'_{i_2}[n'] \leq \Delta t$, then
\begin{equation} \label{eq:laxity_diff2}
L'_{i_1}[n'+1] - L'_{i_2}[n'+1] \leq \Delta t.
\end{equation}
This is because if $L'_{i_1}[n'] - L'_{i_2}[n'] \leq 0$, since $L'_{i_1}[n' + 1] \leq L'_{i_1}[n'] + \Delta t$ and $L'_{i_2}[n'+1] \geq L'_{i_2}[n']$, then \eqref{eq:laxity_diff2} follows.

Otherwise, $0 < L'_{i_1}[n'] - L'_{i_2}[n'] \leq \Delta t$, and the BS will allocate more resource to user $i_2$ than $i_1$. Hence, $L'_{i_1}[n'+1] - L'_{i_1}[n'] \leq L'_{i_2}[n'+1] - L'_{i_2}[n']$ and $L'_{i_1}[n'+1] - L'_{i_2}[n'+1] \leq L'_{i_1}[n']- L'_{i_2}[n'] \leq \Delta t$.

By the definition of ULT, we know that $i_1 \leq_n i_2$ implies the existence of an $n'$ ($0\leq n' \leq n$), such that $L'_{i_1}[n']  - L'_{i_2}[n'] \leq 0 ~(< \Delta t)$. Thus, the desired results follows.

%%%%%%%%%%
\section{Proof of Theorem \ref{thm:asymp_opt_LLHPR_diff_arr}}\label{app:proof_of_asymp_opt_LLHPR_diff_arr}
%%%%%%%%%%

Without loss of generality, we assume that $A_1 = 0$ and $D > 0$. For given time $t > 0$, let $i_{\text{last}}(t)$ be the last user enters the system before time $t$, i.e.,
\begin{equation*} \label{eq:}
i_{\text{last}}(t) = \max\{i: 1 \leq i \leq M, A_i < t\}. \nonumber
\end{equation*}
We divide the interval $[0,t]$ into $N$ time slots, each of which with length $\Delta t = t /N$. Assume that the $i$-th ($1 \leq i \leq i_{\text{last}}(t)$) user arrives in the $n_i$-th slot, i.e., in the interval $(n_i \Delta t, (n_i+1)\Delta t]$, and can be served from time slot $n_i + 1$. The initial virtual expected laxity of user $i$ is $L'_i[n_i+1] = L'_i(A_i)= D - \frac{F_{i,0}}{g_1}$. We study the lower bound on the least virtual expected laxity of L$^2$HPR by examining the least-laxity-set $\underline{\mathcal{Q}}[N]$.

Similar to the proof of Lemma \ref{thm:laxity_bound}, we know that for all $i \in \underline{\mathcal{Q}}[N]$,
\begin{equation} \label{eq:bound_laxity_diff_het_arr}
L'_i[N] \leq \underline{L}_{\text{L$^2$HPR}}[N] +  (M-1)\Delta t.
\end{equation}

Assume that the users in the least-laxity-set $\underline{\mathcal{Q}}[N] = \{i_{1}, i_{2}, \ldots, i_{\underline{Q}[N]}\}$ are sorted in the ascending order of their arrival times. If some of the users in $\underline{\mathcal{Q}}[N]$ is completed, then the sum of virtual expected laxity in time slot $N$ is bounded by
\begin{equation} \label{eq:}
\underline{L}_{\text{L$^2$HPR}}[N] \geq D - (M-1)\Delta t.
\end{equation}

Otherwise, by the definition of $\underline{\mathcal{Q}}[N]$, we know that from time slot $n_{i_{j}} +1$ to $n_{i_{j+1}}$ ($j = 1, 2, \ldots, \underline{Q}[N] -1$), all the $j$ largest data rates $\{g_1, g_2 - g_1, \ldots, g_{j} - g_{j-1}\}$ are allocated to the users in $\underline{\mathcal{Q}}[N]$, and from time slot $n_{i_{\underline{Q}[N]}} +1$ to $N-1$,  all the $\underline{Q}[N]$ largest data rates $\{g_1, g_2 - g_1, \ldots, g_{\underline{Q}[N]} - g_{\underline{Q}[N]-1}\}$ are allocated to the users in $\underline{\mathcal{Q}}[N]$. Thus, the sum of virtual expected laxity is
\begin{eqnarray} \label{eq:}
\sum_{i\in \underline{\mathcal{Q}}[N]} L'_i[N] &=& \sum_{i\in \underline{\mathcal{Q}}[N]} L'_i[n_i+1] + \Delta t \sum_{j = 1}^{\underline{Q}[N]-1}g_j(n_{i_{j+1}} - n_{i_{j}}) \nonumber \\
&& + \Delta t g_{\underline{Q}[N]}(N - n_{i_{\underline{Q}[N]}}-1).
\end{eqnarray}

Then, as $N$ tends to infinity, the slot length $\Delta t = t /N$ tends to 0. From \eqref{eq:bound_laxity_diff_het_arr}, we know that the virtual expected laxity of all users in $\underline{\mathcal{Q}}(t)$ tends to the least virtual expected laxity at time $t$, i.e., $\underline{L}_{\text{L$^2$HPR}}(t)$. Therefore, we have $\underline{L}_{\text{L$^2$HPR}}(t) = D$ when all the users are complected, or
\begin{eqnarray} \label{eq:}
\underline{L}_{\text{L$^2$HPR}}(t) &=& \frac{1}{\underline{Q}(t)}\bigg\{\sum_{i\in \underline{\mathcal{Q}}(t)} L'_i(A_i) + \sum_{j = 1}^{\underline{Q}(t)-1}\frac{g_j}{g_1}(A_{i_{j+1}} - A_{i_j}) \nonumber \\
&& + \frac{g_{\underline{Q}(t)}}{g_1}(t-A_{i_{\underline{Q}(t)}})\bigg\},
\end{eqnarray}
which is the maximum least-virtual-expected-laxity can be obtained by any scheduling policies. Thus, for any schedulable arrival sequence, the L$^2$HPR policy reaches the maximum least-virtual-expected-laxity at any time $t$, and is asymptotically optimal when $\Delta t \to 0$.

\bibliography{PDNetOpt}

\end{document}